\documentclass[twocolumn,prd,amssymb,amsmath,preprintnumbers,floatfix,aps]{revtex4-1}
\usepackage{dcolumn,graphicx,bm,hyperref}
\newcommand{\be}{\begin{equation}}
\newcommand{\ee}{\end{equation}}
\newcommand{\bear}{\begin{eqnarray}}
\newcommand{\eear}{\end{eqnarray}}
\newcommand{\ba}{\begin{array}}
\newcommand{\ea}{\end{array}}
\begin{document}
\preprint{Fermilab-Pub-19-252-T}

\title{\boldmath \bf \Large General solution to the $U(1)$ anomaly equations  \vspace{0.2cm}}

\author{\bf Davi B.~Costa$^\diamond$, Bogdan A.~Dobrescu$^\star$ and Patrick J.~Fox$^\star$  \vspace{0.3cm} }   

\affiliation{$\diamond$ Universidade de  S\~ao Paulo, 
SP 05508-090, Brasil  \looseness=-1  \vspace{0.08cm}  
 \\
$\star$ Theoretical Physics Department, Fermilab, Batavia, Illinois, USA  \looseness=-1 \vspace{0.2cm} } 

\date{\normalsize  May 31, 2019  \vspace{0.3cm} }

 \vspace{0.3cm}

\begin{abstract}
The  anomaly cancellation equations for the $U(1)$ gauge group can be written as a 
cubic equation in $n-1$ 
integer variables, where $n$ is the number of Weyl fermions carrying the $U(1)$  charge.
We solve this Diophantine cubic equation by providing a parametrization of the charges
in terms of $n-2$ integers, and prove that this is the most general solution.
\end{abstract}

\maketitle

Gauge symmetries are essential to the remarkable success of the Standard Model of particle physics in describing 
all measured properties of the known elementary particles.
In order to be well-behaved at high energies, gauge theories must be free of gauge anomalies, including 
the ones generated by fermion loops with three gauge bosons on the external lines \cite{Bardeen:1969md}. 
Furthermore,  in a fundamental theory, the  
elementary fermions are expected to be chiral, {\it i.e.}, their charges should not allow 
a mass term larger than the scale of spontaneous symmetry breaking.
This requires careful assignment of the fermion charges under the gauge group. 

$U(1)$ gauge symmetries acting on chiral fermions are present in  many theories of interest for new  physics, 
as well as within the Standard Model which includes hypercharge. 
The triangle anomaly with three $U(1)$ gauge bosons on the external lines, {\it i.e.}, the $[U(1)]^3$ anomaly, is cancelled provided a cubic equation with fermion charges 
as variables is satisfied. 
The triangle anomaly with one $U(1)$ gauge boson and two gravitons on the external lines should also be cancelled 
\cite{Eguchi:1980jx}. This gives an additional anomaly equation, which is linear in fermion charges.
In the presence of additional gauge groups, there are more equations that need to be satisfied, but 
the cubic and linear ones are always necessary.

Any $U(1)$ gauge group must be embedded in a non-Abelian structure in order to ensure that its gauge boson is well behaved at high energies.
As a result, it is generically expected that the $U(1)$  charges must be commensurate,   {\it i.e.}, they can be restricted to have integer values for certain
normalizations of the gauge coupling. Thus, the $U(1)$  anomaly cancellation conditions are equivalent to Diophantine equations, 
{\it i.e.}, equations with integer variables. 
Finding anomaly-free sets of fermions is challenging, as there are no generic methods of solving Diophantine cubic equations  
\cite{Hardy}.  

Any chiral set of fermions can be embedded in a larger set of fermions which is chiral and anomaly-free  \cite{Batra:2005rh}.
However, there is less known about how to identify anomaly-free sets with a fixed number of fermions charged under a  $U(1)$ gauge group.
Numerical methods are useful, but become difficult to use for a large number of fermions or for large ratios of charges \cite{Allanach:2018vjg}.
Methods based on algebraic geometry \cite{Rathsman:2019wyk} are of limited scope so far.
Deriving anomaly-free sets from the $U(1)$ subgroups of a non-Abelian gauge group, such as $SO(10)$ or $E_6$, is a powerful technique
\cite{deGouvea:2015pea}, yet it is restricted by  the choice of the larger group. 
Examples of chiral $U(1)$ models can be found in  \cite{Nakayama:2011dj}.

In this letter we solve the cubic and linear anomaly equations for the $U(1)$ gauge group in the most general way.
Our main observation that leads to a general solution is that an anomaly-free set of chiral charges can be constructed out
of two vectorlike sets.
For a set of $n$ chiral fermions we identify a parametrization of the charges in terms of $n-2$ integers, and then we prove that 
any solution of the anomaly equations corresponds to certain values for these integer parameters.

\subsection*{Anomaly-free chiral sets   for the $U(1)$ gauge group} 
 
\vspace*{-2mm} 
 
Consider a number, $n$, of Weyl fermions 
carrying nonzero charges $z_1,  ... , z_n$ under a $U(1)$ gauge group. Without loss of generality the $z_i$'s are taken as integers. There are two anomaly equations:
the $[U(1)]^3$ anomaly cancellation requires a Diophantine cubic equation,
\be
z_1^3 + \cdots + z_n^3 = 0 ~~,
\label{eq:cubic}
\ee
and the linear $U(1)$ anomaly cancels when 
\be
z_1 + \cdots + z_n = 0 ~~.
\label{eq:linear}
\ee

If among the $n$ Weyl fields there is a pair of vectorlike fermions, {\it i.e.}, their charges satisfy $z_i = - z_{i'}$, then that pair 
does not contribute to the anomalies and the problem reduces to $n-2$ fermions.
Thus, it is sufficient to consider chiral sets of fermions, which means that the following condition must be satisfied:
\be
z_i + z_{i'} \neq 0 \;\;\; {\rm for}  \;\;\;  {\rm any}  \;\;\; 1 \leq i,i' \leq n  ~~~.
\label{eq:chiral}
\ee
Anomaly-free chiral sets exist only for $n \ge 5$ fermions \cite{Batra:2005rh}.

For each solution to the anomaly equations (\ref{eq:cubic}) and (\ref{eq:linear}), a second solution can be obtained by flipping the sign of all charges. 
Furthermore, since  the equations are symmetric under charge permutations, for any solution there are $n! -\!1$ other 
solutions obtained by changing the ordering within the set. 
To eliminate this plurality, we define the {\it canonical form} for a set of charges as the ordering  according to their decreasing absolute values,
making  the first charge  positive.  
Thus, the chiral set $\{ \vec{z} \, \} \equiv \{z_1, ... , z_n \}   $, with  $z_i \in \mathbb{Z} $ for $1 \leq i \leq n$,
is in the canonical form if it satisfies
\be
z_1 \ge |z_2| \ge ... \ge |z_n| \ge 1    ~~~.      
\label{eq:canonical}
\ee

As Eqs.~(\ref{eq:cubic}) and (\ref{eq:linear}) are homogeneous, for any chiral set 
which satisfies them there is an infinite set of solutions where all the charges are multiplied by an
arbitrary integer. It is therefore sufficient to consider coprime sets of charges,  {\it i.e.},  sets where 
the greatest common divisor (gcd) of the $n$ charges is 1. 
For large enough $n$ the solutions include ``composite" sets, formed of two or more 
subsets which independently satisfy the anomaly equations. 
As these are easier to construct, we will focus on chiral  sets of charges which are coprime and non-composite.
We refer to this type of chiral sets as {\it primitive solutions} to the anomaly  equations.

Our goal is to find the most general solution for any $n$. Given that there are two equations, 
finding the most general solution entails identifying
a parametrization of the $n$ charges in terms of at most $n-2$ integer parameters, and then proving that any primitive solution 
can be obtained for certain values of the parameters.
To grasp the difficulty of the problem, extract $z_n$ from Eq.~(\ref{eq:linear}), and write  Eq.~(\ref{eq:cubic}) as the Diophantine equation
\be
z_1^3 + \cdots + z_{n-1}^3 = \left( z_1 + \cdots + z_{n-1}   \right)^3   ~~.
\label{eq:cubic2}
\ee
This is a homogenous cubic equation collectively in the $n-1$ variables,
but also a quadratic equation in  each variable.
The equation in $z_{n-1}$  has a discriminant $\Delta$ given by a quartic polynomial in other charges. 
The question becomes for what  $z_i \in \mathbb{Z}$  with $i \leq n-2$  is $z_{n-1} $ an integer.
Even the prerequisite problem, of when is $\sqrt{\Delta}$ a rational number, is highly nontrivial.

\subsection*{A construction method for anomaly-free sets}  

\vspace*{-2mm} 

We now introduce a method of generating anomaly-free sets under the $U(1)$ gauge symmetry. 
Our main observation is that given two 
integer solutions to Eqs.~(\ref{eq:cubic}) and (\ref{eq:linear}), $\{\vec{x}  \, \} \equiv \{ x_1,  ... , x_n\}$ and $\{\vec{y}  \,  \} \equiv \{ y_1,  ... , y_n\}$, 
another integer solution can be constructed by taking a certain linear combination of the two sets with coefficients which are cubic polynomials in the charges:
\be
\hspace*{-0.2cm}
\{ \vec{x}  \,  \}   \oplus   \{\vec{y}  \,  \}   \equiv   \left(  \sum_{i=1}^n x_i y_i^2 \right)  \{ \vec{x}  \,  \}  -  \left(  \sum_{i=1}^n x_i^2  y_i \right)   \{\vec{y}  \,  \}  ~~.
\label{eq:xyzset}
\ee
We will refer to the operation, labelled by $\oplus$,  that acts on the $ \{ \vec{x}  \,  \} $  and  $ \{\vec{y}  \,  \} $ 
sets as the `merger', and to its result as the merged set. The merger operation satisfies
\bear
&  \{ \vec{y}  \,  \}   \oplus   \{\vec{x}  \,  \} = -  \{ \vec{x}  \,  \}   \oplus   \{\vec{y}  \,  \}   ~~,
  \nonumber \\ [-1.3mm]
\label{eq:properties}
 \\ [-1.3mm]
&   \{ - \vec{x}  \,  \}   \oplus   \{\vec{y}  \,  \} =   \{  \vec{x}  \,  \}   \oplus   \{  - \vec{y}  \,  \}  = \{  \vec{x}  \,  \}   \oplus   \{\vec{y}  \,  \}   ~~.
  \nonumber
  \eear
The merger can be applied again on $\{ \vec{x} \, \} \oplus   \{ \vec{y} \, \} $ and either $\{ \vec{x} \, \} $ or $ \{ \vec{y} \, \} $, and the same procedure can be 
repeated, leading  to sequences of parametric solutions to the anomaly equations.

There is no guarantee that the merged  set  is chiral. 
If one or both of $\{\vec{x}  \, \} $ and $\{\vec{y}  \, \}$ are chiral, the merger $\{ \vec{x} \, \} \oplus   \{ \vec{y} \, \} $ is sometimes vectorlike. 
More surprisingly, the merger of two vectorlike sets is often chiral. We use this observation to construct solutions to the anomaly equations.

\subsection*{Solution for an even number of fermions}  

\vspace*{-2mm} 

As the vectorlike sets with an even  number $n$ of Weyl fermions 
 look qualitatively different than the ones with odd $n$, we treat the two cases separately. Our main result for 
even $n$  is that a chiral solution is generated by the merger of the following vectorlike sets:
\bear
&&    \hspace*{-0.2cm}   \{  \vec{v}_+ \} = \{   \ell_1, k_1, ... \, , k_m , -  \ell_1 , -k_1, ... \, , - k_m \}      ~~,
\nonumber \\ [-0.6mm]
\label{eq:evenv1v2}
 \\ [-0.8mm]
&&    \hspace*{-0.2cm}   \{ \vec{v}_- \} = \{  0, 0,  \ell_1, ... \, ,  \ell_{m} , -  \ell_1, ... \, , -  \ell_{m} \}  ~~,
\nonumber
\eear
where $m = n/2 - 1 \geq 2$, and the $n-2$ parameters   ($k_i$, $\ell_i$, $1\leq i \leq m$) 
are  integers. Note that $\ell_1$ is the only parameter common to both sets.
Since $ \{  \vec{v}_+ \} $ and  $\{  \vec{v}_- \} $ are vectorlike sets, they are anomaly free so their merger, 
\be
\{ \vec{z}\, \} =  \{  \vec{v}_+ \}   \oplus      \{  \vec{v}_- \}   ~~,
\label{eq:even}
\ee
is automatically a solution to 
the anomaly equations (\ref{eq:cubic}) and (\ref{eq:linear}).
Before  proving that this solution
is  the most general one (up to an integer rescaling), we discuss the constraints on its parameters.

The merger operation is a vector sum with coefficients given by
cubic polynomials in charges, introduced in Eq.~(\ref{eq:xyzset}), which in this case are 
\bear
 \hspace*{-0.4cm}   S_+ &\!=&  
  \sum_{i=1}^{m-1}  ( k_{i+1} \!-\! k_{i}) \, \ell_{i}^2     -  \,  (\ell_1 \! + \! k_m)  \, \ell_{m}^2  ~~,
\nonumber \\ [-1.3mm]
\label{eq:SxSy}
 \\ [-3.3mm]
  \hspace*{-0.4cm}  S_- &\!=&   k_1^2  \ell_1  +  \sum_{i=2}^{m}  k_{i}^2 \, (  \ell_{i}\! -\! \ell_{i -1} )  -   \ell_1^2 \ell_{m}    ~~.
\nonumber
\eear
Explicitly, the charges of the merged set are  given by 
\bear
\{ \vec{z}\, \} & \!= \! &  \left\{  \ell_1 S_+ \, , \,  k_1 S_+ \, , \, k_2 S_+ \! + \ell_1 S_-  , \,  ... \, ,  k_m S_+ \! +  \ell_{m-1} S_-   ,    \rule{0mm}{4mm}   \right.
\nonumber \\ [0.1mm]
&&  \left.      \hspace*{-0.6cm} 
 - \ell_1 S_+ \! +  \ell_{m} S_-  , \,   - k_1 S_+ \! - \ell_1 S_-  , \,  ... \, , \,  - k_{m} S_+ \! -  \ell_{m}  S_-  \rule{0mm}{4mm} \right\}  .
 \nonumber \\ [-0.5mm]
  \label{eq:evenset}  
  \eear
The charges of this set are quartic polynomials in the integer parameters.
If one is interested in finding only primitive solutions, then $\{ \vec{z}\, \}$ must be divided by  gcd($\vec{z}\,$)
to ensure that the set is coprime.
Without further restrictions, each solution is generated by several choices of  the  integer parameters. For example, 
flipping the sign of all the parameters leaves the set invariant.

We are interested in a set with $n$ nonzero charges, which implies that certain values of the integer parameters should be
avoided: $\ell_1, k_1 \neq 0$, \ $S_+ \neq 0$, \ $k_2 S_+ \neq - \ell_1 S_-$, {\it etc}.
The chirality conditions  (\ref{eq:chiral}) are satisfied provided 
other $n(n-1)/2$ restrictions are imposed on the integer parameters.
The simplest of these are  $ \ell_m \neq 0$,  $S_- \neq 0$, 
$ \ell_1 \neq - k_1$, $k_m  \neq \ell_1$, and   $\ell_{i}  \neq  \ell_{i-1}$, $k_{i} \neq  k_{i-1}$ for $i = 2, ..., m$.
More complicated chirality conditions are of the types $( k_i + k_{i'})S_+  \neq  - (\ell_{i-1} + \ell_{i'-1})S_-$, 
 $ ( k_i \!-\! k_{i'})S_+  \!\neq \! - (\ell_{i} \!-\! \ell_{i'})S_-$, $(\ell_1 - \! k_{i})S_+  \neq  \ell_{i} S_-$, {\it etc}.
 
\subsection*{Proof of generality for even $n$} 

\vspace*{-2mm} 

In order to prove that the most general solution for an even number $n$ of chiral fermions 
is given by the set $\{ \vec{z} \, \}$ of Eq.~(\ref{eq:even}), we consider an arbitrary chiral set $\{ \vec{q}  \, \}$ which is a solution to Eqs.~(\ref{eq:cubic}) and (\ref{eq:linear}), 
and identify $n-2$ integers such that $\{ \vec{z}  \, \}$ is the same as $\{ \vec{q} \, \}$ up to an overall rescaling by an integer.

To do that we first determine ratios of integer parameters in terms of the charges of Eq.~(\ref{eq:evenset}):
\bear
&& \hspace*{-2cm}
 \frac{k_1}{\ell_1}  =  \frac{q_2}{q_1}  
\nonumber \\ [2mm]
&& \hspace*{-2cm}
\frac{1 }{\ell_1}   \left( k_{i+1} - k_i  \right) =  \frac{1}{q_1}    \left(  q_{i +2} +  q_{m+i+2}  \right)  
\nonumber \\ [-5mm]
  \label{eq:ratiokl}
\\ [3mm]
&&  \hspace*{-2cm}
\frac{1}{\ell_1}  \left(  \ell_{i+1} -  \ell_{i} \right)   = \frac{ q_{i +2} +  q_{m+i+3}  }{ q_2+q_{m+3} } 
\nonumber
\eear
for any $1\leq i \leq m - 1$. From these recursive relations one can determine $n-3$ integer parameters as rational functions of the $\{ \vec{q} \, \}$ charges times $\ell_1$.
Many of these rational functions can be simplified by using the linear equation $q_1 + ... + q_n =0$. 
There is one more nontrivial relation that follows from summing over a pair of charges: $S_-/S_+ = - (q_{2} +   q_{m+3})/q_1$.
However, this relation also  follows from Eq.~(\ref{eq:SxSy}) and the anomaly equations for  $\{ \vec{q} \, \}$.  

We need to  solve Eq.~(\ref{eq:ratiokl})   with $k_i,\ell_i$ as integer variables.
The solution up to a common integer  rescaling, for an arbitrary $\{ \vec{q}  \, \}$, is 
\bear 
&& \hspace*{-0.3cm} k_{i} = (q_2+q_{m+3} )  \sum_{\alpha = 2}^{i+1}   \left( q_{\alpha } + (1 \! - \! \delta_{\alpha ,2})  \, q_{m+\alpha } \rule{0mm}{4mm}  \right)  ~,
\nonumber   \\ [0.1mm]
&&  \label{eq:dif}
 \\ [- 5.3mm]
&& \hspace*{-0.3cm} \ell_{i} = q_1    \sum_{\alpha = 3}^{i+2}   \left(q_{\alpha-1} + q_{m+\alpha} \right) ~~,
\nonumber  
\eear
for $1\leq i \leq m$, where $\delta_{\alpha,2}$ is the Kronecker symbol.
We have thus established that for any 
anomaly-free set $\{ \vec{q} \, \}$  there are some integers $k_i,\ell_i$ 
such that  the $\{ \vec{z}  \, \}$ set  introduced in Eqs.~(\ref{eq:evenv1v2}) and (\ref{eq:even}) is proportional to $\{ \vec{q} \, \}$. 
The above expressions for the parameters 
imply that the $\{ \vec{z}  \, \}$ set is given by $\{ \vec{q}  \, \}$ multiplied by an integer:  
\be
\{ \vec{z}  \, \} =  - q_1 S_{_-}     \{ \vec{q}  \, \}   ~~.
\label{eq:rescale}
\ee
The remaining issue is what happens for a $\{ \vec{q}  \, \}$  that gives $S_- =0$. We point out that under 
the $q_2 \leftrightarrow q_{m+3}$ interchange only $k_1$ is modified. Given that $\{ \vec{q}  \, \}$  is chiral,
Eq.~(\ref{eq:dif}) implies $\ell_1 \neq 0$, so that $S_-$ changes when $k_1$ changes. Thus, the  $S_- =0$  case 
 is avoided by changing the ordering within $\{ \vec{q}  \, \}$.

This completes the proof that any solution   $\{ \vec{q}  \, \}$ can be generated, up to an integer rescaling,  
by a certain choice for the integer parameters in our set of Eq.~(\ref{eq:even}).

\begin{table}[t]
\begin{center}
\renewcommand{\arraystretch}{1.5}
\begin{tabular}{|c|c|c|}\hline  
 \hspace*{0.1mm} Primitive solution $\{ \vec{z} \, \}$/gcd$(\vec{z} \, )$  \hspace*{0.1mm}
 & \ \  $(k_1,k_2,\ell_1, \ell_2)$  \  \ &    \hspace*{0.5mm}gcd$(\vec{z} \, )$\hspace*{0.2mm}  
\\ \hline \hline
    $ \{ 5, -4, -4, 1, 1, 1 \} $   &  $( 1, -2, 1,2 )$    &  1
\\ [1mm]    \hline
      $  \{ 6, -5, -5, 3, 2, -1  \}    $    &  $( 2, 0, 1, -1 )$    &  1
\\ [1mm]    \hline
     $  \{   11, -9, -9, 4, 4, -1\}   $     &  $( 2, 3, 2, -2  )$    &  8
\\ [1mm]    \hline
      $   \{  11, -9, -9, 5, 1, 1 \}   $     &  $(1, 3, 1, -1  )$    &  2
  \\ [1mm]    \hline
     $  \{  11, -10, -8, 5, 4, -2 \}   $     &  $( -1, 2, 2,  -1   )$    &  2
\\ [1mm]    \hline
 \ \   $   \{  12, -11, -10, 8, 6, -5 \}   $  \  \ &  $(  3, 2, 2, 3  )$    &  10
\\ [1mm]     \hline
\end{tabular}
\vspace{-0.1cm}
\caption{Primitive solutions to the anomaly equations for 6  fermions
 in canonical form  for $z_1 \leq 12$. A   choice of the $k_1$, $k_2,\ell_1, \ell_2$ 
  parameters  and the greatest common divisor that generate the primitive solution from the 
set (\ref{eq:6set}) are shown. 
\\ [-9mm] }
\label{table:1}
\end{center}
\end{table}

\subsection*{ Anomaly-free chiral sets  with 6 Weyl fermions}  

\vspace*{-2mm} 

Let us consider the particular case of $n=6$ fermions. The cubic equation (\ref{eq:cubic})  takes 
the symmetric form 
\be 
 \hspace*{-0.1cm} 
(z_1 \!+\! z_2) (z_2 \!+\!  z_3) (z_3\!+\!  z_1) = - (z_4 \!+\!  z_5) (z_5 \!+\!  z_6) (z_6 \!+\!  z_4)  ~.
\label{eq:cubic6}
\ee
The general solution  (\ref{eq:even})  for $n =6$  is parametrized  by 4 integers, $k_1,k_2,\ell_1, \ell_2$:
\be
\{ \vec{z}\, \} =  \{  \ell_1, k_1, k_2,  - \ell_1, -k_1, - k_2 \}  \oplus  \{  0, 0, \ell_1,  \ell_2, - \ell_1, - \ell_2\}      ~.
 \label{eq:even6}
\ee
Its charges are given by quartic polynomials: 
\bear
&&  \hspace*{-0.3cm} 
z_1 = \ell_1  \left(  \ell_1^2 ( k_2 - k_1 ) - \ell_2^2 (\ell_1 + k_2 )   \rule{0mm}{4mm} \right) 
\nonumber \\ [1.2mm]
&&  \hspace*{-0.3cm} 
z_2 =  k_1 \left(  \ell_1^2  (k_2 - k_1) - \ell_2^2 (\ell_1+k_2 )    \rule{0mm}{4mm} \right)
\nonumber \\ [2mm]
&&  \hspace*{-0.3cm} 
z_3 = \ell_1^2 k_1 ( k_1 - k_2) -  \ell_2 (\ell_1 + k_2) (\ell_1^2 -  \ell_1 k_2 + k_2 \ell_2)
\nonumber \\ [-1.mm]
\label{eq:6set}
\\ [-1.5mm]
&&  \hspace*{-0.3cm} 
z_4 = \ell_2^2 k_2 (\ell_1 + k_2 ) -  \ell_1 (k_2 - k_1 ) (\ell_1^2 + k_1 \ell_2 + k_2 \ell_2)
\nonumber \\ [2mm]
&&  \hspace*{-0.3cm} 
z_5 = \ell_1^2 k_2 (k_2 - k_1 ) +  \ell_2 ( \ell_1 + k_2 ) (\ell_1^2 - \ell_1 k_2 + k_1  \ell_2 )
\nonumber \\ [2mm]
&&  \hspace*{-0.3cm} 
z_6 = \ell_1 \left(    \ell_2 ^2(\ell_1 + k_2 )   +  (k_2 -k_1) ( k_2 \ell_2 - \ell_1 k_2 + k_1 \ell_2 )   \rule{0mm}{4mm} \right)
\nonumber 
\eear
A necessary condition for this set to be chiral is $k_1, \ell_1,  \ell_2 \neq 0$, while   $k_2 =0$ typically gives a chiral set.
Table I collects the primitive solutions in canonical form with $z_1 \leq 12$, a corresponding choice of $(k_1,k_2,\ell_1, \ell_2)$, 
and the gcd($ \vec{z}\,$). 

The canonical form (\ref{eq:canonical}) 
implies that $z_i \!+\! z_{i'}$ has the same sign as  $z_i$ for $i \! < i'$, so Eq.~(\ref{eq:cubic6}) requires $z_2z_5 \! <  \! 0$. 
As Eq.~(\ref{eq:cubic6})  is invariant under charge permutations, the  interchange $z_1 \leftrightarrow z_4$ gives 
$z_3 z_5 < 0$,  and $z_2 \leftrightarrow z_5$ leads to $z_3 z_4 < 0$. 
Then Eq.~(\ref{eq:linear}) allows only two signatures: $\{+,-,-,+,+,\pm \} $.
  
\subsection*{General solution for an odd number of fermions} 

\vspace*{-3mm} 

Our key result for an odd number $n$ of Weyl fermions is that the general solution is given by the merger of the two vectorlike sets
\bear
\{ \vec{u}_+\} &=& \{0, k_1, ... \,  , k_{m+1}, -k_1, ... \,  , -k_{m+1} \} ~~,
\nonumber \\ [-1.1mm]
\label{eq:evenu1u2}
 \\ [-1.6mm]
\{ \vec{u}_-\} &=& \{ \ell_1, ...  \, , \ell_m, k_1, 0, -\ell_1, ... \, , - \ell_m, -k_1 \}   ~,
\nonumber 
\eear
where $m = (n-3)/2 \geq 1$. The $n-2$ integer parameters are $k_1, ... , k_{m+1}, \ell_1 , ... , \ell_m$, 
with only $k_1$ common to both sets.  The merged set  is then  
\bear
\hspace*{-8mm} 
\{\vec{z} \, \} &=&  \{ \vec{u}_+ \}   \oplus      \{ \vec{u}_- \}
 \nonumber\\ [2mm]
\hspace*{-8mm} 
&=& \left\{ \ell_1 S_-  \, , \,   k_1 S_+\!  + \ell_2 S_-   \, , \,   ...   \, , \,  k_{m-1} S_+\!  +  \ell_m S_-  \, ,
 \right. \nonumber\\ 
&&    k_{m}S_+\!  +  k_1 S_-  \, , \,  k_{m+1} S_+   \, , \,  -  k_1 S_+\!  -  \ell_1 S_-  \, , \, \,    ...  \, , \, \,  
\nonumber\\
&&\left.  - k_{m} S_+\!  -  \ell_m S_-  \, , \, \,  -k_{m+1} S_+\!  -  k_1 S_- \right\}~, \label{eq:generaloddmerge}
\eear
with $S_\pm$ given in terms of the $\{\vec{u}_{\pm}\}$ charges  as in Eq. (\ref{eq:xyzset}),    
\bear
&& \hspace*{-0.7cm}  S_+ =  \sum_{i=1}^{m-1}k_i(\ell_{i+1}^2 \!-\ell_{i}^2)  + k_{m}(k_1^2 \! -\ell_m^2)-k_{m+1} k_1^2  ~,
\nonumber\\[-1.7mm]
\label{eq:Spmoddn}
 \\ [-1.7mm]
&&  \hspace*{-0.7cm}  S_- = \sum_{i=1}^{m-1}k_i^2(\ell_{i}-\ell_{i+1}) + k_{m}^2(\ell_{m}- k_1)+k_{m+1}^2 k_1~.\nonumber
\eear
The conditions for nonzero charges and for chirality are $\ell_1, k_1, k_{m+1} \neq 0$,   $S_\pm \neq 0$,  $k_1\neq \ell_m$,    
$\ell_i \neq \ell_{i+1}$,  $k_i \neq k_{i+1}$, and 
more complicated ones, similar to the even $n$ case.

Following the analysis for even $n$, we compare $\{\vec{z} \, \}$ to an arbitrary anomaly-free chiral set 
$\{\vec{q}\,\}$ and determine $k_i,\ell_i$   such that $\{\vec{z}\,\} \! \propto \! \{\vec{q}\,\}$.
From Eq.~(\ref{eq:generaloddmerge}) we   find 
\bear
&& \hspace*{-0.2cm}
\frac{1}{k_1}\left(k_{i+1}-k_{i}\right)= \frac{q_{i+1}+q_{m+i+3}}{q_1+q_{m+3}}   \;\; , \;\;  1\leq i \leq m  ~~,
\nonumber \\ [1mm]
&& \hspace*{-0.2cm}
\frac{\ell_1}{k_1} = -\frac{q_1}{q_{m+2}+q_{n}}    ~~~,
  \label{eq:oddrecurrence}
 \\  [1mm]
&& \hspace*{-0.2cm}
\frac{1}{k_1}\left(\ell_{i+1}-\ell_{i}\right) = -\frac{q_{i+1}+q_{m+i+2}}{q_{m+2}+q_{n}}   \;\; , \;\;  1\leq i \leq m - 1 ~~,
\nonumber 
\eear
and  also  $S_-/S_+ = (q_{m+2}+q_{n})/(q_{1}+q_{m+3})$.  
Using these recursive relations we derive an integer solution for the $k_1, ... , k_{m+1}$ and $\ell_1 , ... , \ell_m$ parameters:
\bear
k_i &=& \left(q_{m+2}+q_{n}\right) \sum_{\alpha=1}^{i}\left(q_{\alpha} + q_{m+2+\alpha} \right)  ~, 
\nonumber\\  [-4mm]
&& 
\\  [-0.2mm]
\ell_i &=& -\left(q_1 \! +q_{m+3}\right)
 \sum_{\alpha=1}^{i} \left(q_{\alpha} \! +  (1 \! - \! \delta_{\alpha,1})  \, q_{m+1+\alpha}   \rule{0mm}{4mm}  \right)  ~~.
 \nonumber
\eear 
This choice for parameters generates the arbitrary set times an integer: $\{\vec{z}\,\} = -  (q_1 + q_{m+3}) S_{_-}  \{\vec{q}\,\}$. 
Again, $S_- \neq 0$ can be avoided by the transposition $q_1 \leftrightarrow q_{m+3}$. 
Combined with the even $n$ case of Eq.~(\ref{eq:rescale}), this shows that our solutions are the most general ones for any $n$.

As an example, note that $n=15$ is the number of Weyl fermions in one generation of Standard Model quarks and leptons.
For $(k_1, ... \,  , k_7)= (-1,1,3,0,-2,0,-2 )$,
 $(\ell_1, ... \,  , \ell_6)=(3,0,3,1,-2,1)$,
and after dividing by gcd=$-2$, Eq.~(\ref{eq:generaloddmerge}) gives the chiral set of charges 
$\{ 6,-4,-4,-4, -3, -3, 2,2,2,1,1,1,1,1,1\} $. 
These are the hypercharges of the quarks and leptons, taken as left-handed fields, 
for a normalization of the $U(1)_Y$ gauge coupling where the quark doublet has hypercharge $+1$.

A nonstandard  example  with  $n=15$ is 
$k_1 = 1$, $\ell_{2 i} = 0$, $\ell_{2 i -1} = k_{2 i + 1} = -k_{2i} = 1$, $i=1,2,3$,
giving the chiral set $\{ 8,8,8,-7,-7,-7,-7,-7,-7,-7,6,6,6,6,1\} $, which is also anomaly free under an $SU(3)\times U(1)$ symmetry.

\subsection*{Conclusions} 

\vspace*{-2mm} 

A long-standing problem in particle physics is how to identify anomaly-free sets of chiral fermions charged under a $U(1)$  gauge symmetry.
This requires solving the anomaly equations, including a Diophantine cubic equation for which there are no generic methods. 
The problem is relevant for dark matter sectors, flavor breaking structures, nonstandard neutrino models, extensions of the electroweak group, 
hidden sectors, and other theories. 

In this article we have solved the  $U(1)$ anomaly equations in general, for any number of chiral fermions, $n$.  
Our solutions, given in Eq.~(\ref{eq:even}) for even $n$, and in Eq.~(\ref{eq:generaloddmerge}) for odd $n$,
are parametrized by $n - 2$ integers.
We have proven that these are the most general solutions, by showing that for any anomaly-free set of charges
there exist $n - 2$ integers which generate that set, 
up to an overall rescaling by an integer (which can be absorbed by a redefinition of the gauge coupling).

To obtain the general solutions, we have developed a method 
of constructing anomaly-free  sets of chiral fermions starting from two 
sets of vectorlike  fermions. The method relies on the ``merger" operation introduced in Eq.~(\ref{eq:xyzset}).
This is a powerful tool to generate chiral $U(1)$ gauge theories with arbitrary fermion content. 

\bigskip
\bigskip

{\it  Acknowledgments:}  {\small   DC is supported by Funda\c{c}\~ao de Amparo \`a Pesquisa do Estado de S\~ao Paulo. 
BD and PF are supported by Fermi Research Alliance, LLC under Contract DE-AC02-07CH11359 with the U.S. Dept. of Energy.
}

\providecommand{\href}[2]{#2}\begingroup

\vfil

\begin{thebibliography}{10}  
  
\bibitem{Bardeen:1969md} 
  S.~L.~Adler,
  ``Axial vector vertex in spinor electrodynamics,''
  Phys.\ Rev.\  {\bf 177}, 2426 (1969). \\
  W.~A.~Bardeen,
  ``Anomalous Ward identities in spinor field theories,''
  Phys.\ Rev.\  {\bf 184}, 1848 (1969). \\
  C.~Bouchiat, J.~Iliopoulos,  
  P.~Meyer,
  ``An anomaly-free  version of Weinberg's model,''
  Phys.\ Lett.\  B{\bf 38}, 519 (1972) \\
  D.~J.~Gross and R.~Jackiw,
  ``Effect of anomalies on quasi- renormalizable theories,''
  Phys.\ Rev.\ D{\bf 6}, 477 (1972).  \\
  H.~Georgi and S.~L.~Glashow,
  ``Gauge theories without anomalies,''
  Phys.\ Rev.\ D {\bf 6}, 429 (1972). 
  
\bibitem{Eguchi:1980jx} 
  For a review, see T.~Eguchi, P.~B.~Gilkey and A.~J.~Hanson,
  ``Gravitation, gauge theories and differential geometry,''
  Phys.\ Rept.\  {\bf 66}, 213 (1980).
  
\bibitem{Hardy}
G.~H.~Hardy and E.~M.~Wright,   ``An introduction to the theory of numbers", 
 Oxford Univ. Press,  6th edition, 2008.
    
\bibitem{Batra:2005rh} 
  P.~Batra, B.~A.~Dobrescu and D.~Spivak,
  ``Anomaly-free sets of fermions,''
  J.\ Math.\ Phys.\  {\bf 47}, 082301 (2006)
  [hep-ph/0510181].

\bibitem{Allanach:2018vjg} 
  B.~C.~Allanach, J.~Davighi, S.~Melville,
  ``An anomaly-free Atlas: charting the space of flavour-dependent gauged $U(1)$ extensions of the Standard Model,''
  JHEP {\bf 1902}, 082 (2019)
  [arXiv:1812.04602]. 

\bibitem{Rathsman:2019wyk} 
  J.~Rathsman and F.~Tellander,
  ``Anomaly-free model building with algebraic geometry,''
  arXiv:1902.08529. 
  
\bibitem{deGouvea:2015pea} 
  A.~de Gouvea and D.~Hernandez,
  ``New chiral fermions, a new gauge interaction, Dirac neutrinos, and dark matter,''
  JHEP {\bf 1510}, 046 (2015)
  [arXiv:1507.00916].   \\  
  J.~M.~Berryman, A.~de Gouvea, D.~Hernandez, 
   K.~J.~Kelly,
  ``Imperfect mirror copies of the standard model,''
  Phys.\ Rev.\ D {\bf 94}, no. 3, 035009 (2016)
  [arXiv:1605.03610]. 

\bibitem{Nakayama:2011dj} 
K.~Nakayama, F.~Takahashi, 
T.~T.~Yanagida,
  ``Number-Theory Dark Matter,''
  Phys.\ Lett.\ B {\bf 699}, 360 (2011)
  [arXiv:1102.4688]. \\
  A.~Ismail, W.~Y.~Keung, K.~H.~Tsao and J.~Unwin,
 ``Axial vector $Z^\prime$ and anomaly cancellation,''
  Nucl.\ Phys.\ B {\bf 918}, 220 (2017)
  [arXiv:1609.02188].  \\
  B.~Batell,
  ``Dark Discrete Gauge Symmetries,''
  Phys.\ Rev.\ D {\bf 83}, 035006 (2011)
  [arXiv:1007.0045]. \\
  Y.~Cui and F.~D'Eramo,
  ``Surprises from complete vector portal theories: New insights into the dark sector and its interplay with Higgs physics,''
  Phys.\ Rev.\ D {\bf 96}, no. 9, 095006 (2017)
  [arXiv:1705.03897]. \\
L.~M.~Cebola, D.~Emmanuel-Costa, R.~Gonzalez Felipe and C.~Simoes,
  ``Minimal anomaly-free chiral fermion sets and gauge coupling unification,''
  Phys.\ Rev.\ D {\bf 90}, no. 12, 125037 (2014)
  [arXiv:1409.0805].  \\
  M.~C.~Chen, D.~R.~T.~Jones, A.~Rajaraman and H.~B.~Yu,
 ``Fermion mass hierarchy and proton stability from non-anomalous $U(1)_F$ in SUSY SU(5),''
  Phys.\ Rev.\ D {\bf 78}, 015019 (2008)
  [arXiv:0801.0248].
  

\end{thebibliography}
\end{document}